# A Photometric and Dynamic Study of Comet C/2013 A1 (Siding Spring) from Observations at a Heliocentric Distance of ~4.1 AU

Yu. S. Andrienko[a], A. V. Golovin[b], A. V. Ivanova[b], V. N. Reshetnik[a, b],
S. N. Kolesnik[a], and S. A. Borisenko[b]

[a]National Taras Shevchenko University of Kyiv, Kyiv, Ukraine
[b]Main Astronomical Observatory, National Academy of Sciences of Ukraine, Kyiv, Ukraine
e-mail: andrienko.j@univ.kiev.ua

**Abstract**—An analysis is presented for the photometric data on comet C/2013 A1 (Siding Spring) from observations at a large heliocentric distance (~4.1 AU). Comet C/2013 A1 (Siding Spring) displays intense activity despite the relatively large heliocentric distance. The morphology of the comet's coma is analyzed. The following parameters are measured: the color indices $V$-$R$, the normalized spectral gradient of the reflec- tivity of the comet's dust $S'$, and the dust production rate $Af\rho$. A numerical simulation is performed for the evolution of the comet's orbit after a close encounter with Mars. The most probable values are obtained for the Keplerian orbital elements of the comet over a hundred-year period. The comet's orbit remains nearly parabolic after passing the orbits of all the Solar System planets.

## INTRODUCTION

The dynamically new comet C/2013 A1 (Siding Spring), which had come from the Oort cloud, was discovered on January 3, 2013 by Robert McNaught at Siding Spring Observatory using the 0.5-meter southern Uppsala Schmidt telescope. At the time of the discovery, at a heliocentric distance of 7.2 AU, the comet displayed photometric activity (McNaught et al., 2013).

Using Hubble Space Telescope (HST) (Li et al., 2014), the comet was observed in three periods: on October 29, 2013, January 21, 2014, and March 11, 2014, at distances of 4.58, 3.77, and 3.28 AU from the Sun, which allowed researchers to investigate the changes in the properties of dust and gas and in the comet's activity with the changing heliocentric distance. A'Hearn et al. (1984) estimated the dust production rate, which was expressed by the value $Af\rho$: 2500, 2100, and 1700 cm for the three distances, respectively. The dust color in the comet's coma for these distances in the spectral range 438–606 nm was found to be $(5.0 \pm 0.3)\%/100$ nm, $(6.0 \pm 0.3)\%/100$ nm, and $(9.0 \pm 0.3)\%/100$ nm, respectively. The reddening was observed to increase with heliocentric distance above ~3000 km from the nucleus.

The comet was observed in the infrared range at a heliocentric distance of 3.82 AU aboard the *NEOWISE* spacecraft on January 16, 2014 (Mainzer et al., 2014). According to model calculations, the dust production rate was about 10 kg/s for grains with a size of about 1.7 and 2.3 μm.

Tricarico et al. (2014) estimated the rate of delivery of dust grains from the comet, using the HST observations. The maximum grain flow estimated from the NEOWISE data was $\sim 10^{-7}$ grains per m$^2$.

On October 19, 2014, comet C/2013 A1 (Siding Spring) made a very close fly-by of Mars. According to recent estimates (Jet Propulsion Laboratory), the minimum distance between the comet and planet during this encounter was $135000 \pm 5000$ km, and their relative velocity was 55.96 km/s (Farnocchia et al., 2014).

By studying the properties of cometary dust at different heliocentric distances (from the regions where the emission of grains from the comet's surface is controlled by highly volatile components (CO and $CO_2$) to the regions of sublimation of water ice at heliocentric distances of less than 3 AU), researchers can study the change in the physical parameters of dust in the different parts of the comet's orbit. The nonstationary processes in comets (emissions associated with the outflow of dust and gas, emergence and disappearance of active regions on comet nuclei and their disintegration, etc.) can lead to qualitative changes in the grain scattering parameters. The spatial distribution of dust grains in the coma is also highly dependent on the level of activity on the surface of the nucleus. The physical characteristics of the dust grains can be determined by analyzing the following photometric parameters of the comet: the color index, excess color, normalized spectral gradient, and change in the surface brightness and dust production.

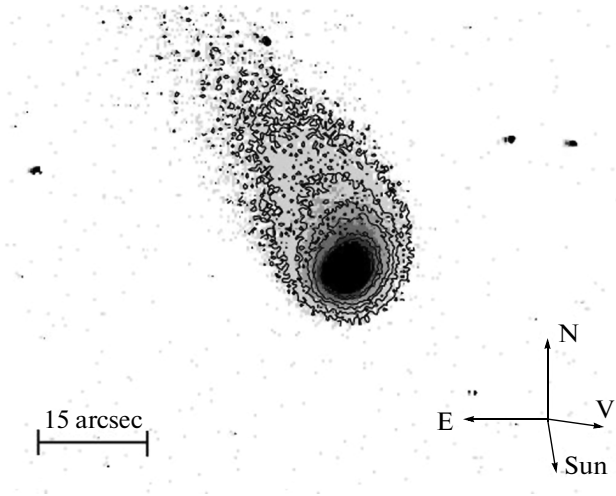

**Fig. 1.** The isophots of comet C/2013 A1 (Siding Spring) for the images of December 12, 2013, in the *R* filter. The figure shows the northward and eastward directions, the direction towards the Sun, and the direction of the comet's motion.

In this paper, we analyze the photometric observations of comet C/2013 A1 (Siding Spring) on December 12 and 13, 2013, when the comet was at distances of 4.161 and 4.151 AU from the Sun, respectively, and study the evolution of the comet's orbit before and after its rendezvous with Mars.

OBSERVATIONS AND PREPROCESSING

The observations of comet C/2013 A1 (Siding Spring) were made using the 2-m Faulkes telescope at Siding Spring Observatory (Australia) on December 12 and 13, 2013, by one of the authors, Aleksandr Golovin. The comet's heliocentric (*r*) and geocentric ($\Delta$) distances as of the said dates were 4.161 AU, 3.645 AU, 4.151 AU, and 3.643 AU, respectively. The comet's phase angle was 12.4°.

A Fairchild CCD camera (486 BIDB) was used as a radiation detector. The image size was 2048 × 2048 px; the scale was 0.3″/px. The CCD field of view was 10.0′ × 10.0′. Images of comet C/2013 A1 (Siding Spring) were obtained using broadband *V* filters (with an effective wavelength of 0.55 µm) and *R* ( with an effective wavelength of 0.65 µm), which sufficiently closely satisfy the Bessel system. All the frames with the comet's images were corrected for the zero point and uneven pixel sensitivity by means of master frames. Since the comet's angular size is much smaller than the size of the frame, we used the standard "sky" procedure of the IDL library to calculate the brightness of the sky background (Landsman, 1993). To increase the signal-to-noise ratio, we added up the comet's images. The final stage of the image processing was to remove the traces of cosmic rays. The quality of the images was measured as the average width of the image profiles for several stars at half the maximum brightness (FWHM), which were obtained in individual images, and was about 1.5″.

An overall view of comet C/2013 A1 (Siding Spring) and a schematic representation of the comet's orbit are shown in Fig. 1. Table 1 presents the journal of observations.

IMAGE ANALYSIS

The comet's images show distinct background stars. We selected all of the bright stars with a signal-to-noise ratio (SNR) of more than 100 as photometric standards. During the calibration, we used 17 stars for the images obtained on December 12, 2013, and 15 stars for those obtained on December 13, 2013. We used standard stars with magnitudes ranging from $10^m$ to $17^m$ from the NOMAD catalogue (http://vizier.u-strasbg.fr/viz-bin/VizieR).

We studied the correlation between the instrumental and catalogue magnitudes and found the coefficients of transformation from the instrumental system to that of *VR* filters from the NOMAD catalogue. During this processing, we did not use the variable stars that fell into the field of view. The comet's integral magnitude as a function of the aperture radius is shown in Fig. 2. The brightness of the background was measured at a distance of 36″–38″ from the comet's center outside the coma and the tail.

**Table 1.** Journal of observations

| Date of observation | Number of expositions | Total exposure time, s | D, AU | r, AU | Phase angle, deg | Air mass | Filter |
|---|---|---|---|---|---|---|---|
| December 12.6283, 2013 | 3 | 180 | 3.645 | 4.161 | 12.3 | 1.15 | V |
| December 12.6329, 2013 | 3 | 180 | 3.645 | 4.161 | 12.3 | 1.17 | R |
| December 13.6273, 2013 | 3 | 180 | 3.643 | 4.151 | 12.4 | 1.16 | V |
| December 13.6359, 2013 | 3 | 180 | 3.643 | 4.151 | 12.4 | 1.21 | R |

To analyze the comet's coma, we built the radial profiles of surface brightness spaced at 1° by the polar angle. The comet center was defined as the maximum of a Gaussian approximating the image of the comet's coma. The resulting profiles were averaged over equal radii near the coma (12″) for all the observations of the same type for every night.

Figures 3 and 4 show the star brightness profile (the point spread function, PSF) and the comet's averaged radial profiles in the $V$ and $R$ filters for December 12 and 13, respectively. The comet's brightness profile is known to consist of two components: the coma and the nucleus. The nucleus profile can be described by the PSF obtained from the field star image.

The changes in the average radial profiles for the two observational nights are shown in Fig. 5 (the solid line shows the profile for December 12 and the dashed line, for December 13). The observations in both filters show an increase in brightness for the central part of the comet (about 2.5″), whereas the outer coma has a constant brightness. The change in the brightness of the central part of the comet with time could indicate an increase in the optical thickness of the nuclear region, which is projected onto the picture plane, which may be a consequence of active processes on the surface of the comet's nucleus and, as a result, the nonuniform structure of the coma.

The resulting radial profiles of the comet's surface brightness were approximated by the power function $B_i(\rho) = A_i \rho^{-\alpha_i}$, where $\rho$ is the cometocentric distance projected onto the picture plane and $i$ is the number of the profile. The parameters $A_i$ and $\alpha_i$ (Fig. 6) were determined for the coma by the least squares method outside the comet nucleus; to exclude the nucleus, we used the value of the instrumental PSF obtained from the field star (Figs. 3 and 4). The dependence of the profile approximation coefficients for the images obtained on December 12, 2013, in the $V$ filter is given in Fig. 6. The azimuth angle is calculated from the direction towards the tail. As is evident from the dependence of the approximating coefficients in Fig. 6, comet C/2013 A1 (Siding Spring) has a small tail in the sector with polar angles of 330°–50°, where the parameters $A$ and $\alpha$ are increasing.

Based on the images obtained in the $V$ and $R$ filters, we determined the integral color indices for comet C/2013 A1 (Siding Spring) in an aperture of radius 9″ $V - R = 0.49 \pm 0.03$ (December 12, 2013) and $V - R = 0.41 \pm 0.02$ (December 13, 2013). The color indices for the comet's coma are greater than for the Sun: $V - R = +0.35$ (Holmberg et al., 2006), which may be evidence of large amounts of dust. The azimuthally averaged profile for the color index is calculated similarly to the radial profiles in the $V$ and $R$ filters and is given in Fig. 7. The solid line corresponds to the observations made on December 12; the dashed one, to those from December 13. It is evident that the comet's

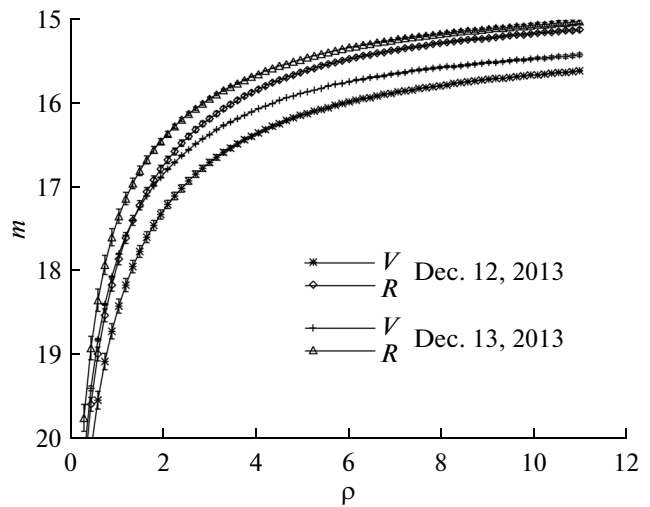

**Fig. 2.** Dependence of the integral stellar magnitudes in the $V$ and $R$ filters on the aperture size in seconds of arc. Here and below, $\rho$ is the angular cometocentric distance in seconds of arc and $m$ is magnitude.

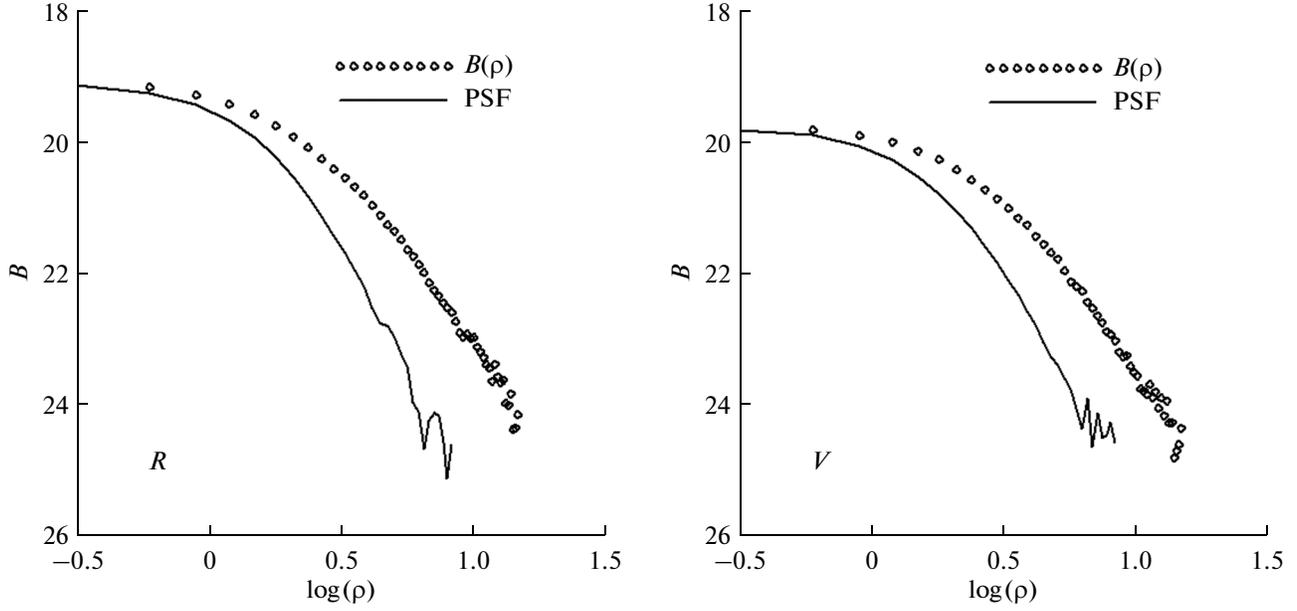

**Fig. 3.** Average radial profiles of the comet's brightness for the images obtained on December 12, 2013. The circle symbols show the comet's brightness $B$ (in magnitudes per square second of arc) with the change in the distance from the central pixel $\rho$ (in seconds of arc) for the $V$ and $R$ filter. The solid curve is the point spread function (PSF).

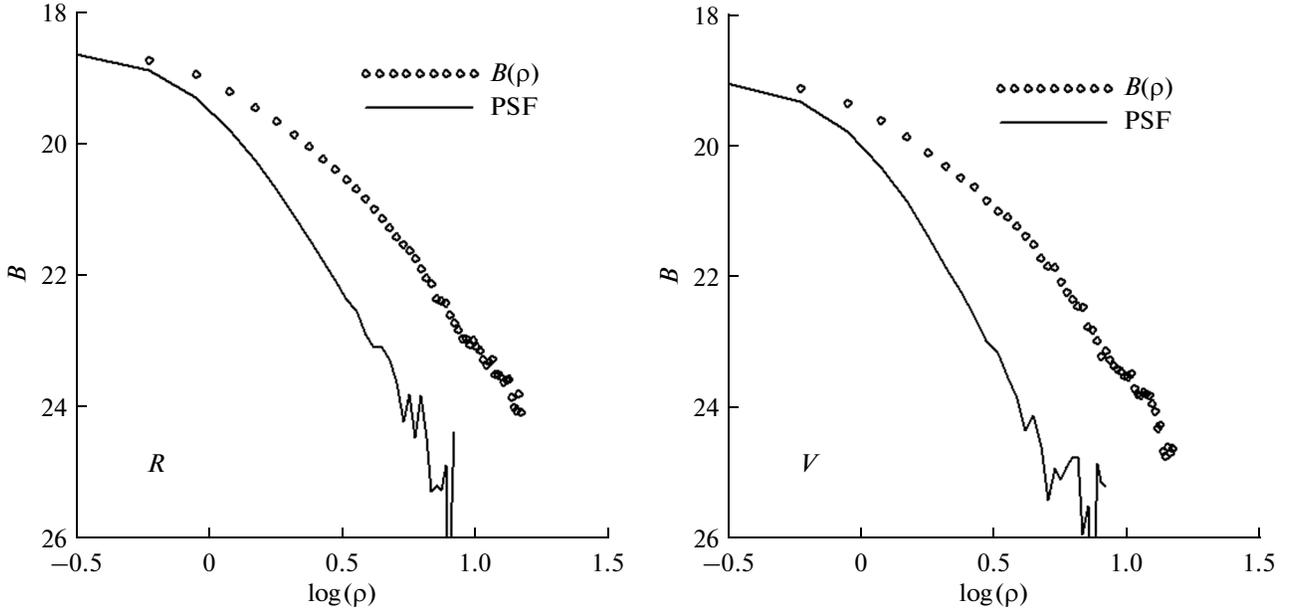

**Fig. 4.** Average radial profiles of the comet's brightness for the images obtained on December 13, 2013. The circle symbols show the comet's brightness $B$ (in magnitudes per square second of arc) with the change in the distance from the central pixel $\rho$ (in seconds of arc) for the $V$ and $R$ filter. The solid curve is the point spread function (PSF).

central part became less red on December 13, compared with the previous day.

The color of a comet is often described parametrically using the normalized spectral gradient $S'$, which is the percentage change of the continuum in the range $10^3$ Å. The percentage of the reddening can estimated by the following formula (A'Hearn et al., 1984):

$$\frac{S'}{100} = \frac{2000}{(\lambda_R - \lambda_V)} \frac{10^{0.4\Delta m_{VR}} - 1}{10^{0.4\Delta m_{VR}} + 1}, \quad (1)$$

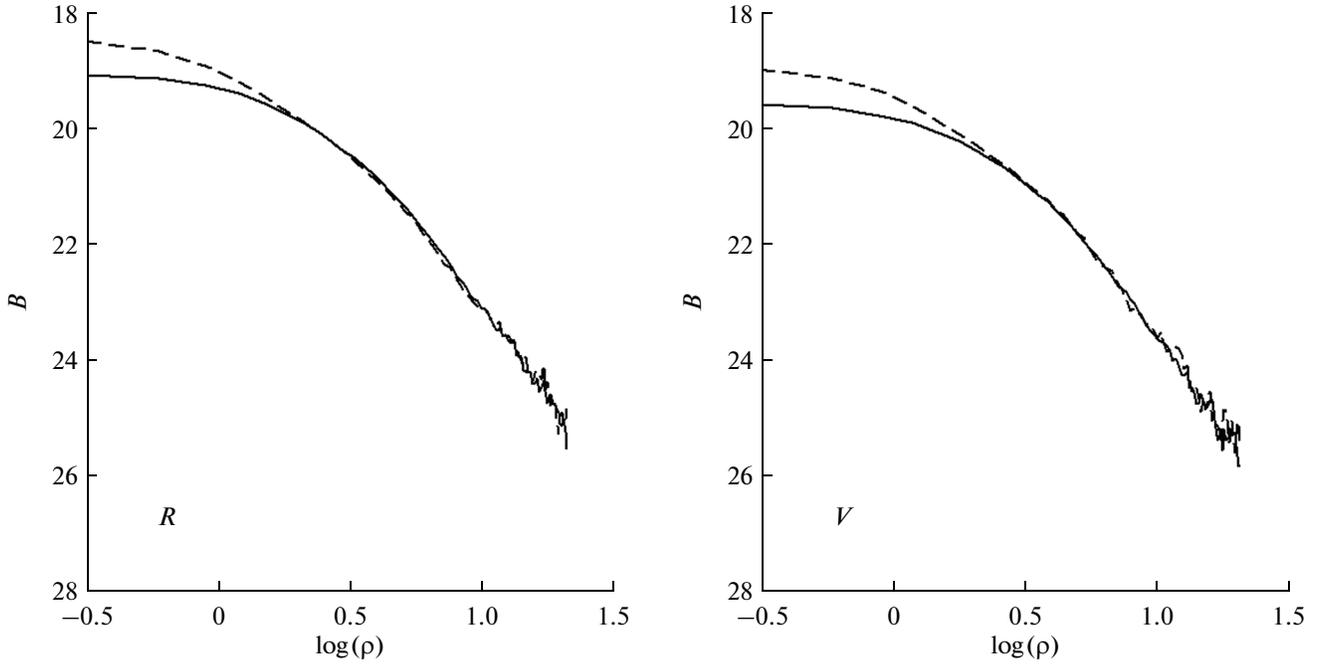

**Fig. 5.** Comet's brightness $B$ (in magnitudes per square second of arc) with the change in the distance $\rho$ from the central pixel (in seconds of arc) for the $V$ and $R$ filters for the images taken on December 12 (solid curve) and 13 (dashed curve), 2013.

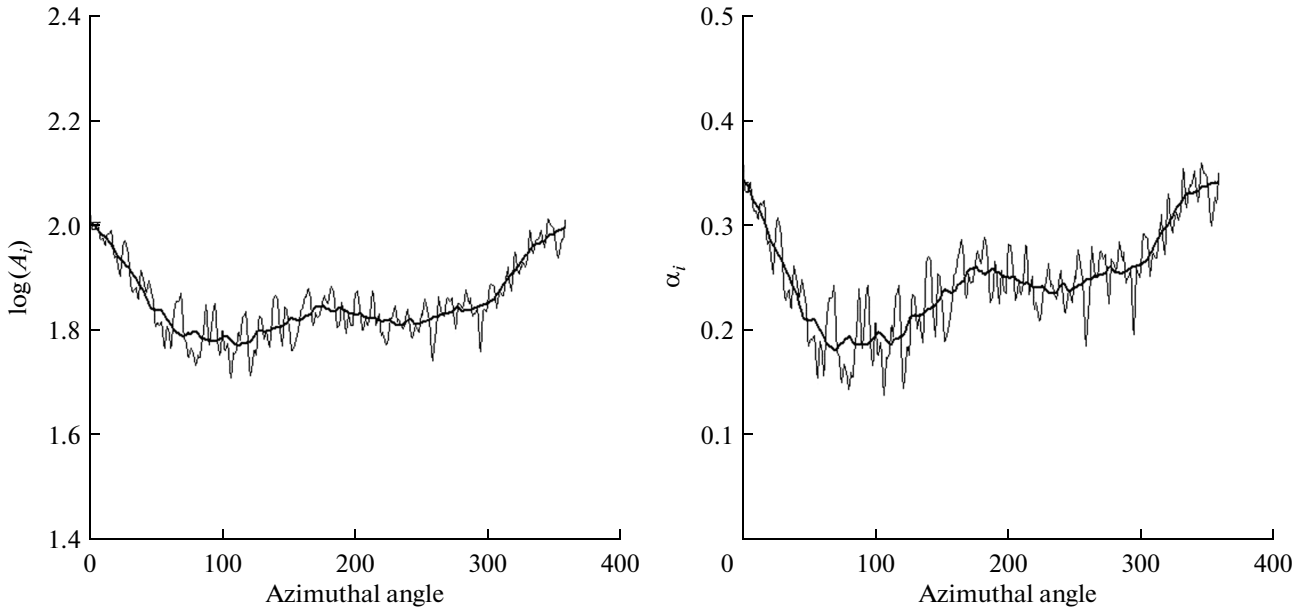

**Fig. 6.** Change in the parameters $A$ and $\alpha$, depending on the azimuthal angles for observations on December 12, 2013. The solid line is the curve smoothed with a rectangular window (with a width of 30°).

where $\Delta m_{VR}$ is the difference between the color indices of the comet and the Sun, or the difference between the effective wavelengths of the applied $V$ and $R$ filters $(\lambda_R - \lambda_V)$ in Å. The dependence of $S'$ on the aperture size is given in Fig. 8.

The analysis of the dust color characteristics for a number of new comets is given in the literature. Newburn and Spinrad (1985) gave a value of $S' \sim 7\%$ at 1000 Å for new comets. Jewitt and Meech (1986) showed that the normalized reflectivity decreases with increasing wavelength; its average value in the optical range is 13%, varying from 5 to 18%. In addition, the measurements for individual comets have a fairly large spread. Thus, Bonev and Jockers (2002) found that the normalized spectral gradient for comet C/2000 WM1 (LINEAR) can be as high as 15%; Korsun et al. (2012)

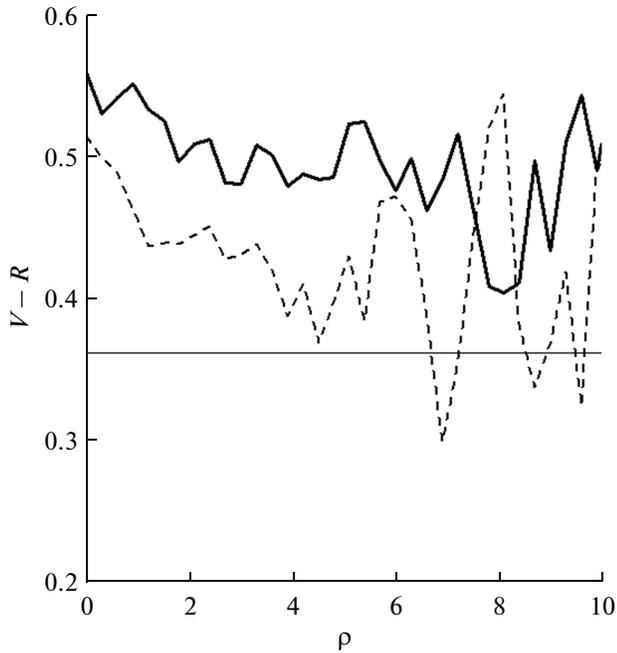

**Fig. 7.** Azimuthally averaged profiles of the color indices of comet C/2013 A1 (Siding Spring) obtained on December 12, 2013, (solid curve) and December 13, 2013, (dashed curve). The horizontal line is the color index of the Sun.

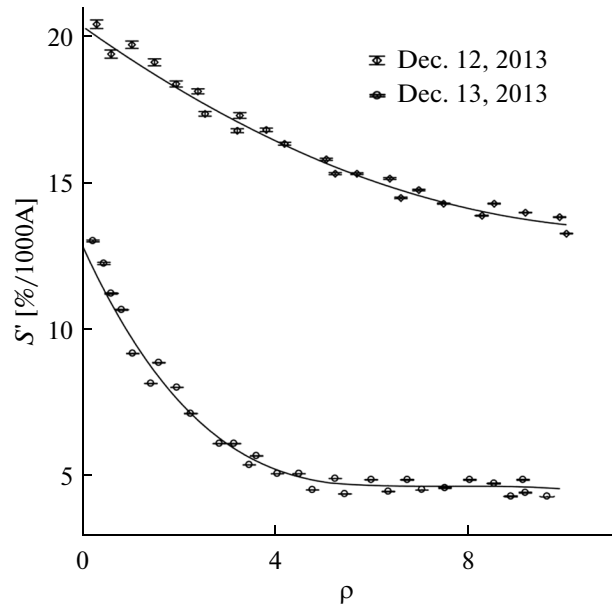

**Fig. 8.** Dependence of the normalized spectral gradient of comet C/2013 A1 (Siding Spring) on the aperture radius ρ. The solid lines are the approximations with second-degree (the curve for December 12, 2013) and third-degree (the curve for December 13, 2013) polynomials.

gave an estimate of 14.3 ± 1.2% for comet C/2010 X1 (Elenin); Borisov et al. (2008) gave an estimate of 6% for comet 8P/Tuttle; and Lin et al. (2007) detected a change in the normalized spectral gradient from 0.6 to 9.0% while observing comet C/2004 Q2 (Machholz).

Our observations indicate changes in the comet's normalized spectral gradient in the wavelength range 550–700 nm over a time period of about one day. These rapid changes are not unique. For example, observations of Halley's Comet also revealed a dependence of the changes in the dust color on the diurnal variations in the comet's activity (Moris and Hanner, 1993).

We also estimated the parameter $Af\rho$, which was derived by A'Hearn et al. (1984). This parameter characterizes the dust production rate of a comet (Korsun et al., 2012) and does not depend on the time and place of observation. The parameter is determined by the ratio of the effective scattering cross-section for all the grains entering the field of view of the detector to the projection of its field of view onto the celestial sphere. The value of $Af\rho$ can be estimated from the magnitude of the dust coma (Mazzotta Epifani et al., 2010):

$$Af\rho = \frac{4r^2\Delta^2 10^{0.4(m_{\text{SUN}} - m_C)}}{\rho}, \qquad (2)$$

where $Af\rho$ is measured in centimeters: $A$ is the average albedo of the dust grains; the filling factor $f$ is the area of the cross-section of all the dust grains in the field of view; ρ is the radius of the field of view in centimeters; $m_C$ is the comet's magnitude; $m_{\text{SUN}}$ is the Sun's magnitude (−26.74 and −27.26 in the $V$ and $R$ filters); $r$ is the comet's heliocentric distance in AU; and Δ is the comet's geocentric distance in centimeters.

Equation (2) can be used to estimate the dust production rate given a uniform outflow of material from the surface. Based on the observations in the $V$ and $R$ filters, we estimated the parameter $Af\rho$ for aperture values from 2″ to 10″ (which corresponds to a distance of $5 \times 10^3$ to $3 \times 10^4$ km); the results are given in Fig. 9.

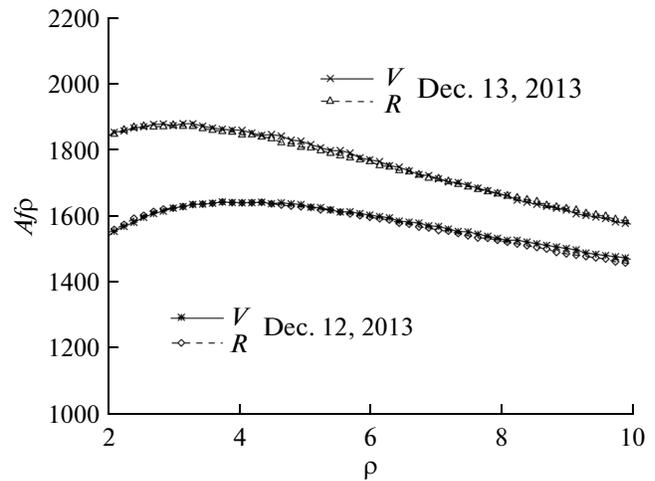

**Fig. 9.** Value of the parameter $Af\rho$ for comet C/2013 A1 (Siding Spring) for different aperture sizes ρ in seconds of arc.

**Table 2.** Accuracy of the determination of the planets' positions over a 100-year period

| Celestial body | Absolute error $\Delta r$, km | Relative error $\delta$ |
|---|---|---|
| Mercury | 0.03 | $5 \times 10^{-10}$ |
| Venus | 0.002 | $2 \times 10^{-11}$ |
| Earth | 0.01 | $6 \times 10^{-10}$ |
| Mars | 0.5 | $2 \times 10^{-9}$ |
| Jupiter | 0.0008 | $8.5 \times 10^{-13}$ |
| Saturn | 0.0002 | $1.5 \times 10^{-13}$ |
| Uranus | 0.0006 | $2 \times 10^{-13}$ |
| Neptune | 0.0003 | $8 \times 10^{-14}$ |
| Pluto | 0.0006 | $8 \times 10^{-14}$ |
| Moon | 5–5.5 | $5 \times 10^{-8}$ |

Figure 9 shows the change in the comet's dust production rate with cometocentric distance and reveals an expansion of the dust coma with the distance from the comet nucleus. It is evident that the slight increase in the dust production rate in the inner part of the coma, which depends on the visibility conditions (image quality), is followed by a decrease in $Af\rho$ with cometocentric distance. This dependence is typical of both of the observation periods. This behavior (the gradual decrease with the distance from the comet nucleus) is often attributed to the failure to satisfy the condition of a perfect equilibrium outflow of material from the comet nucleus (active processes, emissions, etc). In addition, there may be a change in the features of the dust grains, e.g., their disintegration (Lara et al., 2004; Tozzi et al., 2003) and/or an emission of a new amount of dust in the coma (Tozzi et al., 2011). The estimates for $Af\rho$ confirm the hypothesis about the greater activity of the dynamically new comets, especially at large heliocentric distances (Meech, 1988). The value of $Af\rho$ for long-period comets at a distance above 5 AU varies between 38 and 3000 cm (Szabo et al., 2002; Weiler et al., 2003; Tozzi et al., 2003; Mazzotta Epifani et al., 2008; 2009; Meech, 1990; Meech et al., 2009; Ivanova et al., 2014). The results for comet C/2013 A1 are in this range.

## ORBIT SIMULATION FOR COMET C/2013 A1 (SIDING SPRING)

The case of comet C/2013 A1 is used to consider the numerical integration of the equations describing the perturbed motion of test bodies with the initial coordinates and velocities close to cometary ones over a 100-year period and show the effects of the close encounter with Mars on the distribution of the comet's orbital elements at the end of this period.

We conducted the numerical simulation using our own integrator for the differential equations of motion of the Solar System bodies, which is based on a simplified mathematical model of the HORIZONS electronic ephemeris system (http://ilrs.gsfc.nasa.gov/docs/2014/196C.pdf). Currently, the integrator is under continuous testing; the results are compared with the HORIZONS data. The criterion for the comparison is the difference between the radius vectors of the $i$th body in our model and in the HORIZONS model in a barycentric ecliptic coordinate system. The integrator was written according to a predictor–corrector procedure. The explicit Adams procedures are used as a predictor; the implicit ones, as a corrector (Bordovitsyna, 1984). We used the Adams methods of the 12th order and an alternating time step of integration.

Over a 100-year integration period, we can guarantee the absolute ($\Delta r$) and relative ($\delta$) errors given in Table 2 for the positions of the planets, the Moon, and Pluto.

**Table 3.** Keplerian elements of the orbit of comet C/2013 A1

| Orbital element | Designation | Dimension | Value |
|---|---|---|---|
| Eccentricity | $e$ | Dimensionless | 0.9999458751571103 |
| Semimajor axis | $a$ | AU, astronomical unit | 2.590376582621090E+04 |
| Mean anomaly | $M$ | Degrees | 3.599987198535640E+02 |
| Longitude of the ascending node | $\Omega$ | Degrees | 3.009725616379619E+02 |
| Argument of perihelion | $\omega$ | Degrees | 2.410668702492957E+00 |
| Orbit inclination | $i$ | Degrees | 1.290062869558766E+02 |

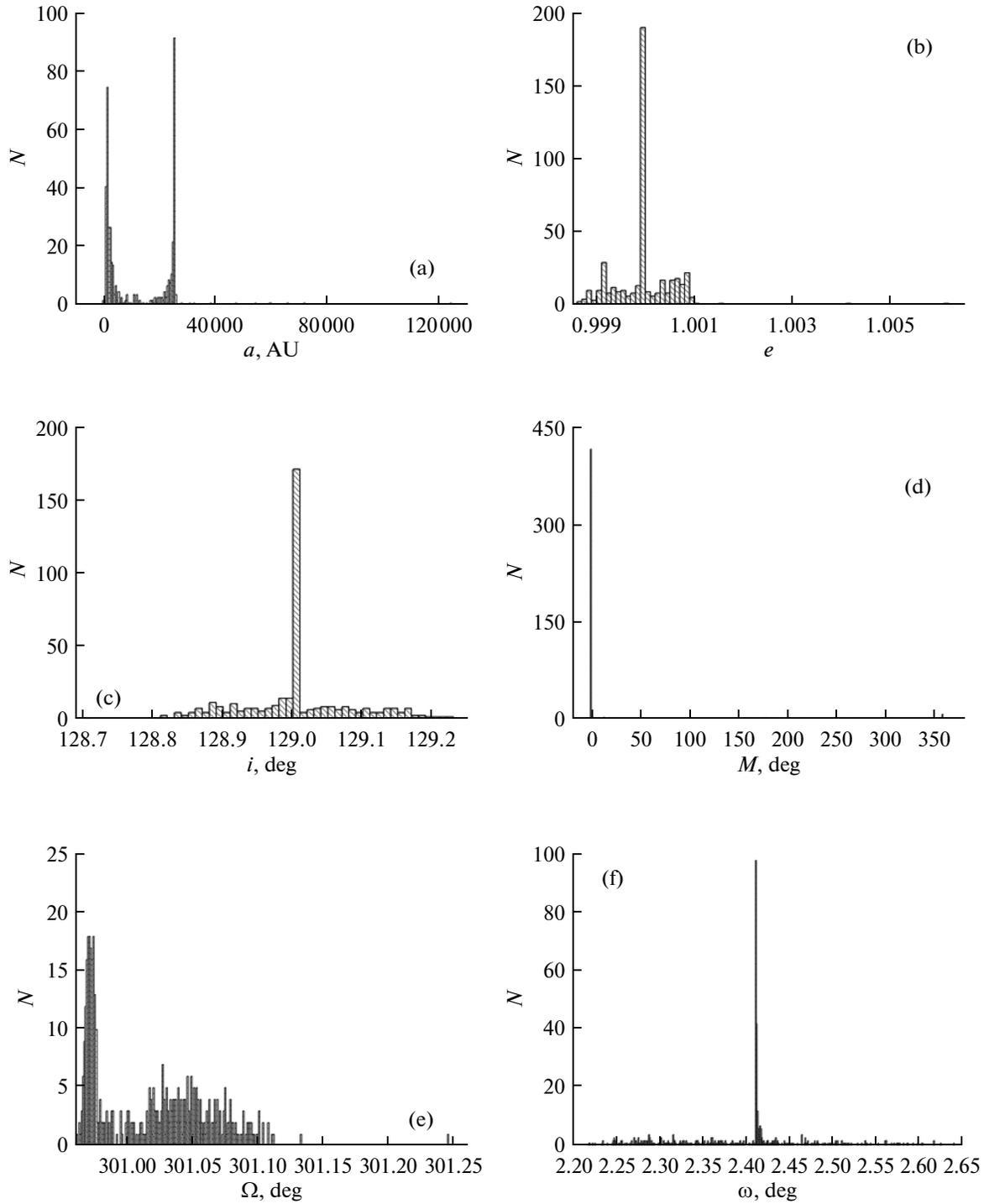

**Fig. 10.** Distribution of the orbital elements of the test cometary bodies at the end of the integration period. (a) Semimajor axis (a); (b) eccentricity (e); (c) orbit inclination (*i*); (d) mean anomaly (*M*); (e) longitude of the ascending node (Ω); (f) argument of pericenter (ω).

The errors for the Moon are much higher than those for the planets because the model does not take into account the time variability of the Moon's inertia tensor, as a result of which the zonal harmonic coefficients $J_2$, $C_{21}$, $C_{22}$, $S_{21}$, and $S_{22}$ in the decomposition of the Moon's potential will also be functions of time. Furthermore, the model does not take into account the perturbations of the Moon's orbit due to terrestrial tides and some other effects described in the available documents for the HORIZONS systems. The main characteristics of our model are as follows:

(1) The Einstein–Infeld–Hoffmann equation is used as the main equation of motion.

(2) The model takes into account the gravitational interactions with the Sun, planets, Moon, Pluto, and 16 of the most massive asteroids and dwarf planets.

(3) The model takes into account the nonspherical shape of the Sun, Earth, and Moon.

(4) The model does not take into account the jet effects on the comet's orbit associated with the gases evaporating from the surface of the comet nucleus.

The initial positions and velocities of all the Solar System bodies included into the integration are taken from the HORIZONS system as of January 1, 2000, 00:00:00 UT. The period of integration is 100 years (January 1, 2000 to January 1, 2100). The orbital positions for all the test cometary bodies are defined by a set of Keplerian elements.

The set of the initial Keplerian elements for comet C/2013 A1 (Siding Spring) in the barycentric ecliptic coordinate system is shown in Table 3.

An important element of the numerical integration is the error in the initial data; even very small uncertainties may prove to have a substantial effect on the long-term prediction of the object's orbit. Therefore, we cloned from the orbit described by a set of Keplerian elements as of the date of integration 443 sets of orbital elements for the test bodies. All the orbital element sets of the test bodies obey the normal distribution for each element. The mean square deviation $\sigma_i$ for each element is of the order of $10^{-6}$ (which is approximately equal to the uncertainty of the Keplerian elements in the HORIZONS system for comet C/2013 A1); the mathematical expectation $\mu_i$ is equal to the initial value of the corresponding orbital element. After the numerical integration of all the orbits of the test bodies, we built distributions of their Keplerian orbital elements as of 2100 (Fig. 10).

The analysis of the distributions suggests that the most probable values of the Keplerian orbital elements of comet C/2013 A1 (Siding Spring) are achieved after passing the Kuiper Belt. The comet's orbit remains nearly parabolic. It is possible to uniquely identify the most probable values for only four out of the six orbital elements (eccentricity, inclination, mean anomaly, and argument of pericenter). As to the semimajor axis $a$, there are two values of this element, which are observed with almost the same probability: 1500–2000 AU and 25 000 AU. The longitude of the orbit's ascending node $\Omega$ has a fuzzy peak of the most probable value at 300.97°.

The preliminary integration of the comet's orbit, which was performed in April 2014 using the initial Keplerian elements derived from a 465-day observation arc, showed that the effects of the comet's nongravitational parameters can be neglected. The resulting error due to the inaccuracy of the known Keplerian orbital elements is much higher than the error due to the neglect of the nongravitational effects. Using the data on the comet's orbit that were obtained from observations in April 2014, we derived the parameters of a close encounter with Mars that are well consistent with NASA's data based on an observation arc of 739 days (http://ssd.jpl.nasa.gov/sbdb.cgi?sstr=C%2F2013%20A1;orb=0;cov=0;log=0;cad=1#cad): the calculated distance of the closest approach is 135 000–137 000 km; the data of the approach is October 19, 2014, 18:27.

## CONCLUSIONS

We analyzed the photometric observations of the dynamically new comet C/2013 A1 (Siding Spring), which had shown a high level of activity at heliocentric distances of about 4.1 AU. The analysis of the comet morphology gave evidence for the presence of a coma and a dust tail at this distance.

Dynamically new comets are, as a rule, brighter and more active than short-period ones at the same heliocentric distances and have stronger emission in the continuum (Meech et al., 2009; Korsun et al., 2010; Ivanova et al., 2014). As shown by the photometry of comet C/2013 A1 (Siding Spring) in the broadband $V$ and $R$ filters, the comet dust has a larger color index compared with the solar one. A significant decrease in the color index during the day may be due to the rotation and nonstationary activity of the nucleus.

The normalized spectral gradient $S'$ obtained for comet C/2013 A1 (Siding Spring) from the images in a 9″ aperture is (Fig. 8): 13.2 ± 0.03% at 1000 Å for the observations on December 12, 2013; 5.1 ± 0.65% at 1000 Å for the observations on December 13, 2013; and an average of 8% for 1000 Å for the nuclear region (this value is close to that of the Oort cloud objects (Jewitt, 2002)). As with the color indices, we observed variations in the normalized spectral gradient at the different observational nights, which may be due to changes in the comet's activity during this period.

The analysis of the comet's dust production rate confirmed its high activity. The parameter $Af\rho$ has a value of about 1550 cm on December 12, 2013, and 1750 cm on December 13, 2013, (within a radius of 2″) in the $R$ filter. Our results are well consistent with the estimates by Li et al. (2014). Thus, the analysis of the observations on the two nights revealed a possible increase in the comet's activity. These results support the hypothesis about the higher activity of dynamically new comets.

We conducted the integration of the orbit of comet C/2013 A1 (Siding Spring) within the known uncertainty of its orbital elements. The resulting orbital elements allow us to predict the most probable position of the comet during the 21st century, which can be used for photometric and spectroscopic observations of the comet at large heliocentric distances when it is moving away from the Sun. The integration of the already known segments of the orbit near the perihelion showed that the effect of nongravitational forces is negligible. Although this effect reduces the prediction accuracy, it can be ignored in a rapid assessment of the orbital elements of comets that have perihelion distances of more than 1–2 AU.

ACKNOWLEDGMENTS

Observations of comet C/2013 A1 (Siding Spring) were made using a 2-m telescope at Siding Spring Observatory (Australia) with the support of the Faulkes Telescope Project. The authors would like to thank the project for the images of the comet.